\documentclass[10pt,letterpaper]{article}
\usepackage[top=0.85in,left=2.75in,footskip=0.75in]{geometry}

\usepackage{changepage}
\usepackage[utf8]{inputenc}
\usepackage{textcomp,marvosym}
\usepackage{fixltx2e}
\usepackage{amsmath,amssymb}
\usepackage{cite}
\usepackage{nameref,hyperref}
\usepackage{microtype}
\DisableLigatures[f]{encoding = *, family = * }
\usepackage{rotating}
\raggedright
\usepackage[aboveskip=1pt,labelfont=bf,labelsep=period,justification=raggedright,singlelinecheck=off]{caption}
\makeatletter
\renewcommand{\@biblabel}[1]{\quad#1.}
\makeatother

\date{}

\usepackage{lastpage,fancyhdr,graphicx}
\usepackage{epstopdf}
\fancyheadoffset[L]{2.25in}
\fancyfootoffset[L]{2.25in}
\newcommand{\be}{\begin{eqnarray}}
\newcommand{\ee}{\end{eqnarray}}
\begin{document}
\vspace*{0.35in}
% Please capitalize all terms in the title except conjunctions, prepositions, and articles.
\begin{flushleft}
{\Large
\textbf\newline{Strong Selection Significantly Increases Epistatic Interactions in the Long-Term Evolution of a Protein}
}
\newline
\\
Aditi Gupta\textsuperscript{1,2,\S},
Christoph Adami\textsuperscript{1,2,3,*}
\\
\bigskip
\bf{1} Department of Microbiology and Molecular Genetics, Michigan State University, East Lansing, Michigan, United States of America
\\
\bf{2} BEACON Center for the Study of Evolution in Action, Michigan State University, East Lansing, Michigan, United States of America
\\
\bf{3} Department of Physics and Astronomy, Michigan State University, East Lansing, Michigan, United States of America
\\
\bigskip
\S Current Address: Center for Emerging Pathogens at New Jersey Medical School, Rutgers University, Newark, New Jersey,  United States of America\\
% Use the asterisk to denote corresponding authorship and provide email address in note below.
* adami@msu.edu

\end{flushleft}
\section*{Abstract}
Epistatic interactions between residues determine a protein's adaptability and shape its evolutionary trajectory. When a protein experiences a changed environment, it is under strong selection to find a peak in the new fitness landscape. It has been shown that strong selection increases epistatic interactions as well as the ruggedness of the fitness landscape, but little is known about how the epistatic interactions change under selection in the long-term evolution of a protein.  Here we analyze the evolution of epistasis in the protease of the human immunodeficiency virus type 1 (HIV-1) using protease sequences collected for almost a decade from both treated and untreated patients, to understand how epistasis changes and how those changes impact the long-term evolvability of a protein. We use an information-theoretic proxy for epistasis that quantifies the co-variation between sites, and show that positive information is a necessary (but not sufficient) condition that detects epistasis in most cases. We analyze the ``fossils" of the evolutionary trajectories of the protein contained in the sequence data, and show that epistasis continues to enrich under strong selection, but not for proteins whose environment is unchanged.  The increase in epistasis compensates for the information loss due to sequence variability brought about by treatment, and facilitates adaptation in the increasingly rugged fitness landscape of treatment. While epistasis is thought to enhance evolvability via valley-crossing early-on in adaptation, it can hinder adaptation later when the landscape has turned rugged.  However, we find no evidence that the HIV-1 protease has reached its potential for evolution after 9 years of adapting to a drug environment that itself is constantly changing. We suggest that the mechanism of encoding new information into pairwise interactions is central to protein evolution not just in HIV-1 protease, but for any protein adapting to a changing environment. %296 words

% Please keep the Author Summary between 150 and 200 words
\section*{Author Summary}
Evolution is often viewed as a process that occurs ``mutation by mutation", suggesting that the effect of each mutation is independent of that of others. However, in reality the effect of a mutation often depends on the context of other mutations, a dependence known as ``epistasis". Even though epistasis can constrain protein evolution, it is actually very common. Such interactions are particularly pervasive in proteins that evolve resistance to a drug via mutations that create defects, and that must be repaired with compensatory mutations. We study how epistasis between protein residues evolves over time in a new and changing environment, and compare these findings to protein evolution in a constant environment. We analyze the sequences of the human immunodeficiency virus type 1 (HIV-1) protease enzyme collected over a period of 9 years from patients treated with anti-viral drugs (as well as from patients that went untreated), and find that epistasis between residues continues to increase as more potent anti-viral drugs enter the market, while epistasis is unchanging in the proteins exposed to a constant environment. Yet, the proteins adapting to the changing landscape do not appear to be constrained by the epistatic interactions and continue to manage to evade new drugs. %(189 words)

%\linenumbers

\section*{Introduction}

The interactions between the residues within a single protein (within-protein epistasis~\cite{deVisseretal2011}) often determine the structure and function of the protein~\cite{Ortlundetal2007,Tokurikietal2008,SoskineTawfik2010}. Understanding these epistatic interactions is important because they shape the protein fitness landscape and thus guide the evolution of a protein given its genetic background~\cite{Bershteinetal2006,Banketal2015}. At the same time, the environment influences the fitness effects of mutations and their epistatic interactions, and a change in the environment changes the topography of the fitness landscape as well as the epistatic effect of mutations~\cite{HaydenWagner2012}. A typical example of a changing environment for a protein is the introduction of drugs to counteract a pathogen. In that case, a drug-resistance mutation that is beneficial in a drug environment might have a significant fitness cost in the absence of drugs~\cite{Maisnier-PatinAndersson2004,Flynnetal2013}. Even in the drug environment, resistance mutations can incur fitness costs that are mitigated by compensatory mutations, and these fitness-restoring mutations are selected together with the resistance mutations if the fitness of the resulting protein exceeds that of the wild-type in the drug-environment~\cite{MartinezPicado2008}. Other compensatory mutations act by  
re-stabilizing the protein and preventing misfolding or proteolysis~\cite{Bloometal2005,ChangTorbett2011,Arayaetal2012}, while sometimes resistance mutations partly compensate each other's deleterious effects~\cite{Trindadeetal2009}. Moreover, compensatory mutations can appear before resistance mutations if they are not deleterious on their own, paving the way for resistance mutations to appear without incurring significant fitness costs~\cite{HarmsThornton2014,Wellneretal2013,Lunzeretal2010}. Epistatic interactions have been implicated in the evolution of drug resistance in many pathogens, including influenza, malaria, {\it Escherichia coli}, tuberculosis, and {\it Chlamydomonas reinhardtii}~\cite{Lozovskyetal2009, Trindadeetal2009, Borrelletal2013, Lagatoretal2014}.

To understand the long-term evolution of a protein in a dynamic environment in which the selection pressure is persistent, evidence of epistatic interactions at multiple time points is needed as the fitness landscape changes over time with the environment. Previous investigations of the impact of epistasis on evolutionary outcomes analyzed epistatic interactions in the absence or presence of a selection pressure at a single time point, showing that epistatic interactions can facilitate adaptation~\cite{GongBloom2014, Kouyosetal2007, Kouyosetal2012, Hinkleyetal2011,Ostmanetal2012, Kryazhimskiyetal2014, Szameczetal2014, Quandtetal2014}. In particular, computational analyses of HIV fitness landscapes derived from statistical models showed that these landscapes have high neutrality as well as ruggedness, suggesting high potential for epistatic interactions~\cite{Kouyosetal2012}. Further, an analysis of \textit{in vitro} virus fitness measurements from more than 70,000 patients revealed that epistasis explained more than half of the variance in fitness measurements, highlighting the importance of epistasis in the HIV-1 fitness landscape~\cite{Hinkleyetal2011}. 

While these studies emphasize the importance of epistasis in adaptive evolution, here we analyze how epistatic interactions themselves change over time. We focus on the Human Immunodeficiency Virus-1 (HIV-1) protease, a small protein necessary for the production of mature and infectious HIV-1 particles. The HIV-1 protease is a homodimer of two 99 residue chains, and it cleaves the viral polyprotein into active components that are necessary for virus maturation~\cite{BrikWong2003}. Because an inactive protease results in uninfectious viruses, the protease was one of the first targets of anti-retroviral drugs. However, due to its high mutation rate (about 0.3 mutations per genome per generation~\cite{ManskyTemin1995}), HIV-1 quickly evolves resistance to those drugs. These resistance and compensatory mutations, as well as the covariation between residues in HIV-1's protease are well studied~\cite{Brownetal1999, Hoffmanetal2003,Haqetal2009}. 

The evolutionary trajectory of resistance to protease inhibitors is extremely complex. When the virus is first exposed to the changed fitness landscape containing the inhibitor, viral count drops to undetectable levels~\cite{Coffin1995}, as the fitness peak that the viral population occupied in the no-drug fitness landscape is erased. However, some minority variants in the untreated viral population might still possess replicative capacity in the drug environment (albeit much reduced), forming the seeds of drug resistance~\cite{Pennings2012}. 
Indeed, most well-known resistance mutations are associated with significant fitness costs~\cite{Mammanoetal1998,MartinezPicado2008}, so that at the onset of resistance the viral populations are likely still significantly smaller than in untreated individuals. However, mutations that compensate for the fitness defects will readily emerge. While these compensatory mutations can either maintain or reduce viral fitness if they occur on their own, they usually restore the fitness cost of the resistance mutation to some extent, and as a consequence form an epistatic pair with the resistance mutation. Note that the residues that are coupled in the treatment landscape are unlikely to be coupled in the absence of treatment, implying that it is the adaptation to the new landscape that forces the interaction. Once compensatory mutations have restored viral fitness, viral populations return to pre-drug levels and can search for more compensating mutations~\cite{Penningsetal2014}. Exposure to second-line drugs (given to patients that have evolved resistance) repeats the cycle, leading to more resistance mutations~\cite{Morand-Joubertetal2006}, followed by compensatory mutations that are epistatically linked with them. 

As more and more resistance mutations accumulate, loss of thermodynamic stability becomes more and more acute, and compensatory mutations are selected that re-stabilize the protein. While stabilizing mutations are not selected against individually when they occur on their own (compensatory mutations often have a fitness {\em advantage} if they occur in the absence of treatment~\cite{Theysetal2012}), several stabilizing mutations together can be undesirable as they decrease protein variability and therefore can impede evolvability~\cite{TavernaGoldstein2002,BloomArnold2009}. Since epistasis and ruggedness are coupled~\cite{BurchChao1999,KvitekSherlock2011,Ostmanetal2012}, the repeated emergence of resistance mutations and selection for compensatory mutations that interact epistatically with those resistance mutations results in a protein that finds itself in an increasingly rugged fitness landscape.

We hypothesize that a strong selection pressure of treatment increases the epistatic interactions in the long-term evolution of HIV-1 protease, allowing the protein to realize its evolutionary potential. To test this hypothesis, we compare the HIV-1 protease sequences from patients treated with anti-retroviral drugs to protease sequences from untreated patients, collected over a nine year period. This wealth of publicly available data provides an unprecedented opportunity to study the adaptive evolution of a protein in an environment with persistent (perhaps even increasing) selection pressures due to the continued introduction of new drugs as opposed to a protein evolving in the absence of these selection pressures. 

Epistatic interactions between mutations are usually deduced by measuring the fitness effect of the single site mutations as well as the double-mutant. However, obtaining evidence of {\em changes} in epistasis over a long enough period of time is challenging when viral fitness has to be assayed. Here we use the mutual (shared) information between protein loci (residues) as a proxy for epistasis. 
Briefly, information shared between sites is a measure of covariation between the sites, so that knowing the residue at one site makes it possible to predict the residue at the covarying site with accuracy better than chance~\cite{Adami2004}. Thus, information constrains the context of a residue, in the same manner that epistasis constrains the fitness effect of a mutation in the presence of another. 
Using information as a proxy for epistasis is not a new idea (see, e.g.,~\cite{Strelioffetal2010} and references cited therein, as well as~\cite{Anastassiou2007} for an application to gene networks). However, the relationship between epistasis and information is not one-to-one. As a consequence, it is possible that two residues interact epistatically but show no information. Conversely, if two residues have positive information, they {\em must} interact epistatically, at least in a model without spatial effects and infinite population size. Thus, information is a sufficient (but not a necessary) condition for epistasis. We have analyzed the relationship between epistasis and information in random fitness landscapes of two interacting loci, and found that in less than 30\% of cases would a pair of loci be epistatic while showing negligible information (see below). Furthermore, those pairs that showed little information also predominantly showed little epistasis. 

Keeping in mind this limitation, studying pair-wise information (which can be deduced from sequence data) instead of epistasis (which cannot) allows us to use the wealth of time-course sequence data of a protein to study its long-term evolution. Not that residue covariation in the protease can in principle be due to population subdivision rather than epistasis~\cite{daSilva2009}, which would be reflected in a deeply branched phylogenetic structure. However, it turns out that deep branches are rare in HIV-1 phylogeny. Indeed, 
using sequence data from the HIV Stanford Database, Wang and Lee show that amino acid covariation in HIV-1 protease sequences is largely due to selection pressures and not due to background linkage disequilibrium~\cite{WangLee2007}. We corroborate this and find that protease sequences from the same database assume a star-like phylogeny (\nameref{sec:tree}), supporting the notion that any observed residue covariation is largely due to selection pressures and not due to population subdivision and other phylogenetic factors.

We first show how the persistent selective pressure of a drug environment increases residue variability over time, and document changes in physico-chemical properties due to these residue substitutions that suggest non-neutral evolution. 
We then show that epistatic interactions are enriched over time in the protein undergoing continuous adaptive evolution in the presence of drugs, but not in proteins where the environment remained constant. This leads us to conclude that while selection pressures increase per-site residue variability--and thus {\em reduce} information stored at each protein locus on average--the information stored in higher-order interactions increases over time. 

\section*{Results}
\subsection*{High selection pressure increases residue variability at multiple loci in the HIV-1 protease}
To compare the long-term adaptation of the HIV-1 protease in the presence and absence of selection pressure of treatment, we analyzed the protease sequences collected in the years 1998-2006 from patients that did not receive treatment (untreated group), as well as from patients that received treatment (treated group, see Materials and Methods).

\begin{figure}[h] %  
   \centering
  \includegraphics[width=3.0in]{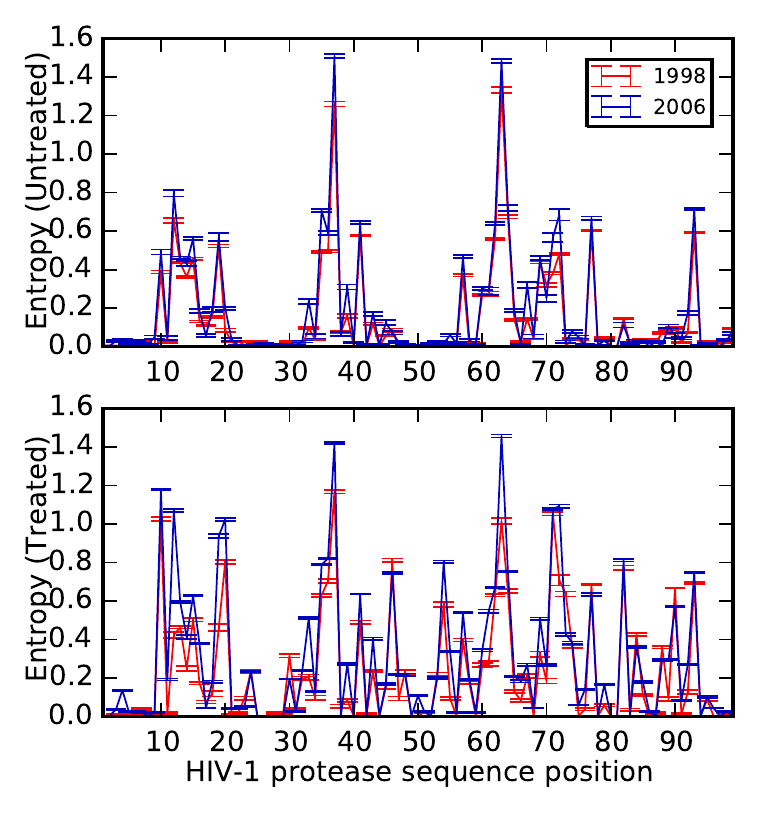} 
   \caption{{\bf Average per-site entropies at every position of the HIV-1 protease.} Untreated (top panel) and treated (bottom panel) datasets at the earliest (year 1998, red) and latest (year 2006, blue) time point of our analysis. 300 sequences are resampled from data for each year and average entropy for each position is calculated from the entropies in 10 resampled datasets. Site-specific variation generally increased across the protein following treatment. Entropy (variability) also increased from 1998 to 2006 for several positions. Error bars denote $\pm 1$ SE.}  
   \label{fig:1998_2006_entropies}
\end{figure}

The per-site entropy (a measure of sequence variation) for each of the 99 positions shows that some protease positions are highly variable even in the absence of treatment, thus indicating some degree of neutrality. However, many more loci become variable (entropic) upon treatment (Fig~\ref{fig:1998_2006_entropies}, see \nameref{sec:logo} for sequence logos of untreated and treated protease sequences averaged over all years). Moreover, several loci in the protein had higher entropy in 2006 compared to 1998, hinting at an increase in per-site residue variability over the years, especially in the treated data set (see also \nameref{sec:deltas}, which shows entropy differences for all years for both the treated and untreated group). 

The increased variability per site might seem counter-intuitive from the point of view of population genetics, where adaptation results in substitutions that lead to reduced diversity from hitchhiking~\cite{Pennings2012}. While such reduced diversity is possible in the short term, we caution that due to the high mutation rate of HIV, pre-treatment diversity can be re-established fairly quickly after drug resistance has emerged. This rebound in virus titer further allows the virus to find mutational paths to resistance. However, it is also possible that some of the observed variability is due to the stochastic mixing (within the database) of many populations that each took diverse paths to resistance. Some positions in the untreated group also show higher residue variability at the later time point, and this is likely due to transmission of drug-resistant virus to untreated individuals via new infections~\cite{Shaferetal2007, Jakobsenetal2010,Yerlyetal1999}. However, this background signal remains low due to the reduced selection pressure in the untreated group.  

%Fig. 1

\subsection*{Increased residue variability due to treatment mirrors physico-chemical changes in the protein}
\begin{figure}[htbp] %  
   \centering
   \includegraphics[width=4.0in]{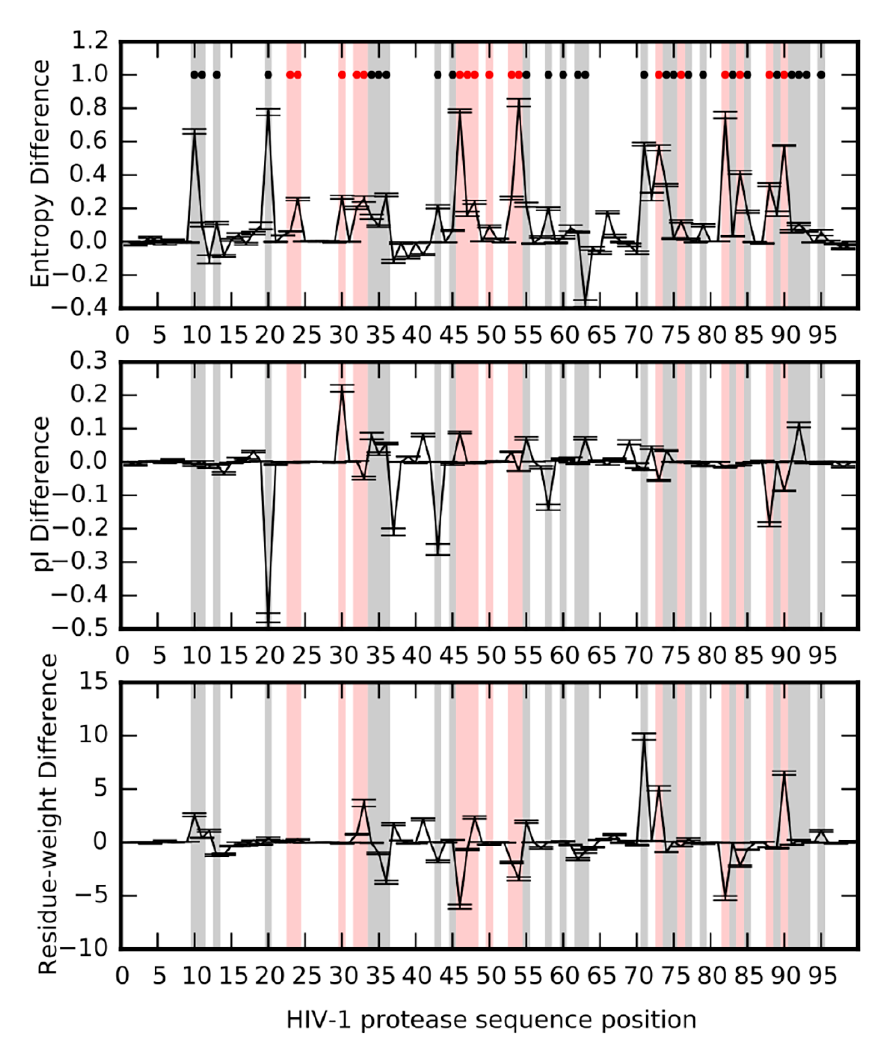} %
      \caption{{\bf Changes in protease entropy and physico-chemical properties.} Changes in per-site entropies (top panel), residue isoelectric point (middle panel), and residue weights (bottom panel) due to treatment. The property difference at each site is obtained by subtracting property (entropy/pI/residue-weight) value of the untreated data from that of the treated data. Average values are obtained by sampling sequence data from all years (1998-2006, 10 subsamples/year of 300 sequences each). Error bars represent $\pm 1$ SE. Red dots represent positions known to be primary drug resistance loci, while black dots mark positions of compensatory or accessory mutations~\cite{ShaferSchapiro2008}. Resistance loci are shaded in red and accessory loci are shaded in black.}
   \label{fig:ent_pI_aawt}
\end{figure}

A closer look at the protease positions with increasing residue variability reveals that the protein loci that have undergone a marked increase in variation have been previously associated with drug resistance (Figure \ref{fig:ent_pI_aawt} top panel)~\cite{ShaferSchapiro2008}. Position 63 is the only site that becomes significantly {\it less} variable upon treatment.  Many of the changes in entropy at each position correlate well with physico-chemical changes (such as changes in the iso-electric point pI or in the residue weight) at those positions (Spearman correlation coefficient between absolute values of entropy difference and pI difference: $R=0.738$, $p=2.95\times 10^{-18}$; Spearman $R$ between absolute values of entropy difference and residue weight difference: $R=0.819$, $p=3.52 \times 10^{-25}$) suggesting that these changes are adaptive and influence the function of the protein in its new environment (see \nameref{sec:chem} and Fig~\ref{fig:ent_pI_aawt}, middle and lower panels). Leucine at position 10 is substituted by the slightly heavier isoleucine, a compensatory mutation showing a slight increase in residue weight difference (Fig~\ref{fig:ent_pI_aawt}, bottom panel, L10I is a compensatory mutation as shown in~\cite{Mammanoetal2000,ShaferSchapiro2008}); lysine is replaced by a more basic arginine residue at position 20, another compensatory mutation as arginine can form more electrostatic interactions compared to lysine and thus enhances protein stability~\cite{Sokalingametal2012,ShaferSchapiro2008}; and aspartic acid is substituted by the uncharged asparagine at position 30, a strong resistance mutation~\cite{Martinez-Picadoetal1999,ShaferSchapiro2008}. While the residue physico-chemical properties discussed here are not exhaustive, the significant positive correlation between changes in these properties and changes in residue variability suggest that the residue variation brought about by treatment is non-random.

%Fig. 2

\subsection*{Epistatic interactions between residues increase following treatment}
Information about a protein's environment is stored not only in individual residues of the protein, but also in the manner in which these residues interact epistatically. We estimate the information the protein stores about its environment in two ways: one that considers explicitly the information stored at each position in the protein independent of other sites ($I_1$), and the measure $I_2$ that in addition to $I_1$ includes the mutual information between every pair of positions (see Materials and Methods).

\begin{figure}[h] % 
   \centering
   \includegraphics[width=4in]{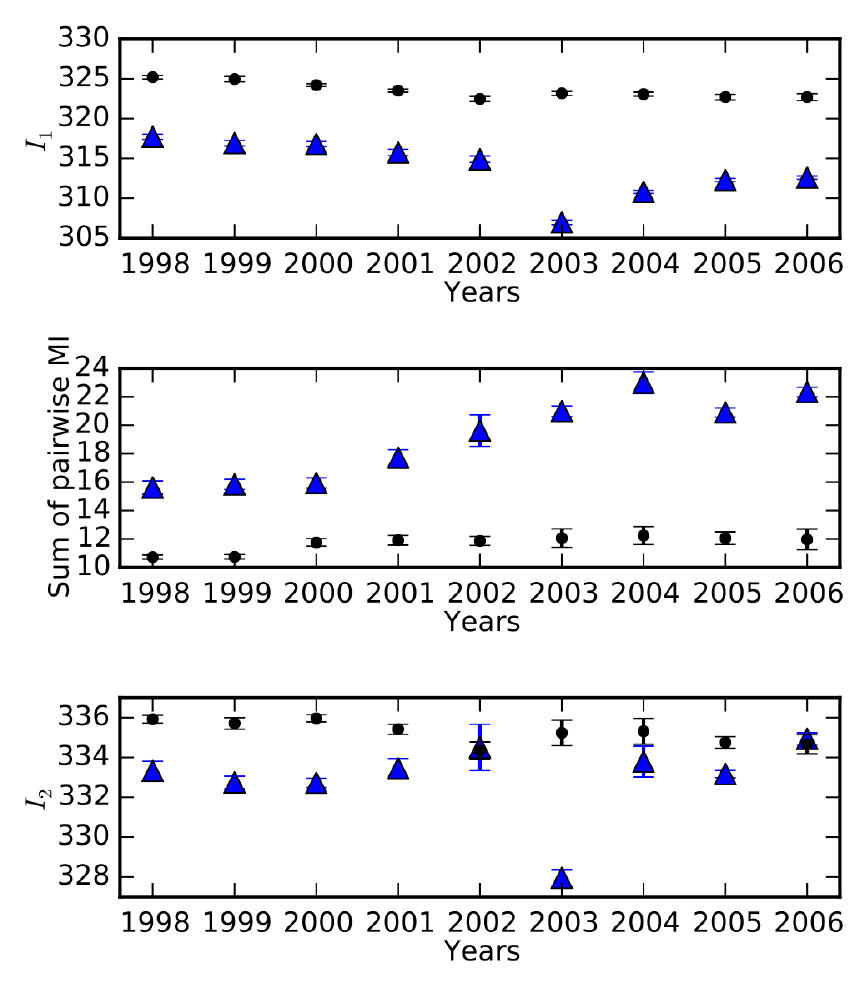} 
   \caption{{\bf Estimates of the information content of the HIV-1 protease}. Filled black circles represent data from untreated subjects and blue triangles represent data from treated individuals.  $I_1$ [see Eq.~(\ref{i1})] is consistently low in treated sequence data over the years, indicating high sequence variability in the drug environment (top panel). The middle panel shows that the sum of pairwise mutual information significantly increases upon treatment ($p \le 0.001$). On adding the sum of pairwise mutual information to $I_1$, we obtain a comprehensive measure of information that considers pairwise interactions between residues [$I_2$, Eq.~(\ref{pairwise})]. $I_2$ for both the treated and untreated data is comparable and unchanging over the years. We use data only for positions 15-90,  as residues 1-14 as well as 91-99 have missing sequence data leading to error-prone estimates of entropy, as evidenced in \nameref{sec:numbers}. Error bars represent $\pm 1$ SD.}
   \label{fig:infoplot} 
\end{figure}

\begin{figure*}[!htbp] %  
   \centering
   \includegraphics[width=5in]{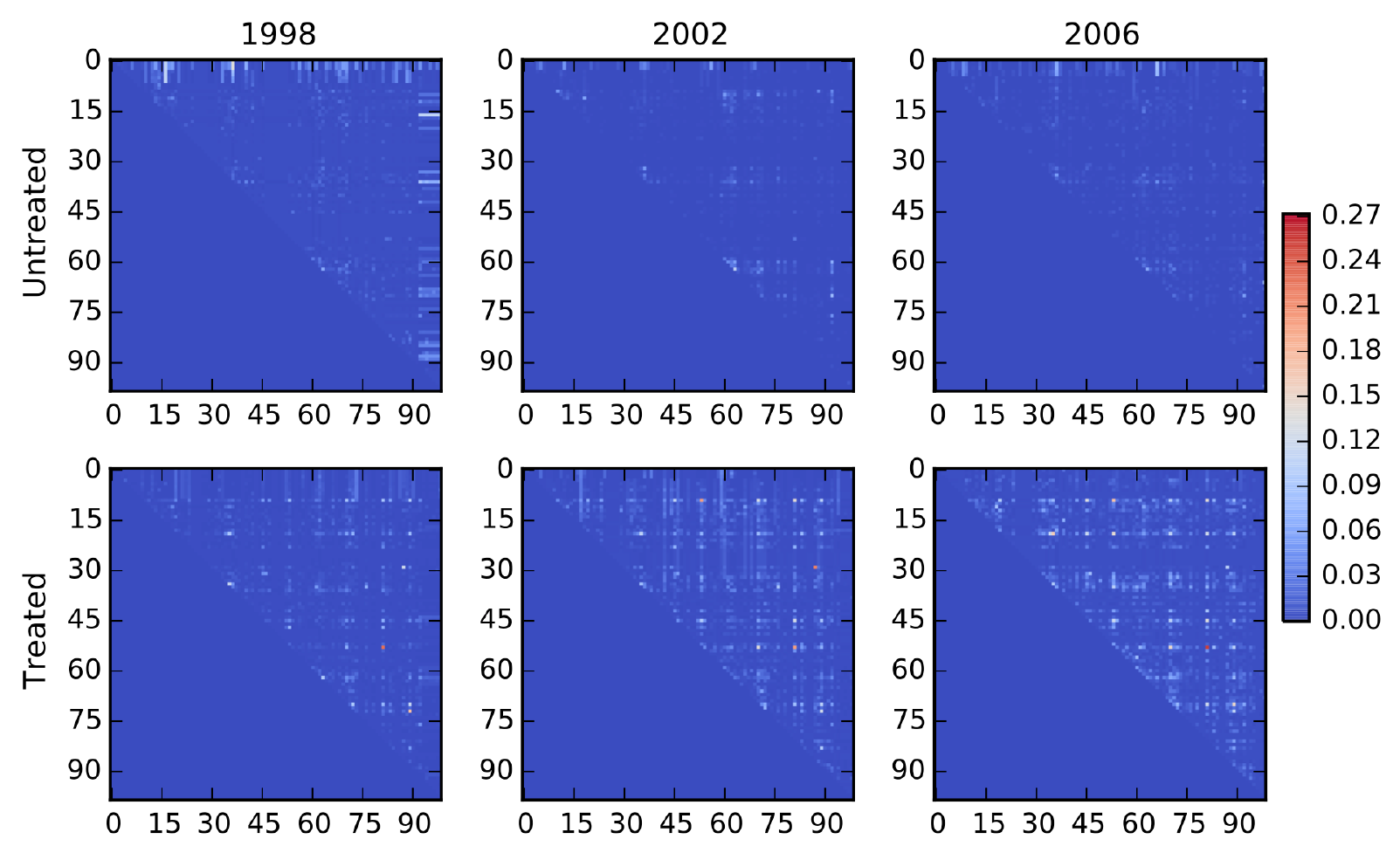} 
   \caption{{\bf Increase in epistasis in HIV-1 protease over time.} Pairwise interactions in the HIV-1 protease are shown for years 1998, 2002, and 2006 in the drug-free (top row) and drug environment (bottom row). Each heatmap shows the mutual information for each pair of residues. Pairwise information (and thus epistatic effects) are fairly constant in the drug-free environment, but gradually increase in the treated group.}
   \label{fig:heatmap}
\end{figure*}

An increase in entropy per site corresponds to a {\rm decrease} in per-site information ($I_1$). We find that the $I_1$ for treated protease sequences is consistently lower that that of untreated sequences, indicating high sequence variability in the treated data (Figure \ref{fig:infoplot}, top panel). The slopes for the treated and untreated $I_1$ data are not significantly different, suggesting that although the treated data has higher sequence variability, this variability does not increase significantly over the years. However, the sum of mutual information of all pairs of positions (the component of information due solely to pair-wise interactions) gradually increases, suggesting that epistatic interactions become enriched over time (Fig~\ref{fig:infoplot}, middle panel, slopes for treated and untreated data are significantly different, $p \le 0.001$). Fig~\ref{fig:heatmap} shows a gradual increase in pairwise mutual information between protease positions due to treatment (bottom panel), while pairwise mutual information in the untreated group remains low over the years (top panel). 
As discussed, adding the sum of pairwise mutual information to $I_1$ gives $I_2$, which thus measures the information stored in the protein taking into account the pairwise dependencies between positions in addition to per-site variability (Fig~\ref{fig:infoplot}, bottom panel, slopes for treated and untreated data are not significantly different). We find that $I_2$ remains relatively constant over the years and converges to the same value in the treated as well as untreated data despite increase in residue variability and enrichment of epistatic interactions in the treated data.  Such a finding may at first appear surprising, but Carothers et al. showed that the functional capacity of biomolecules (RNA aptamers in their study) correlated with the information contained in the sequence, and that functionally equivalent biomolecules had similar total information content~\cite{Carothersetal2004}. Our analysis thus suggests that the similar total information content for proteins in the treated and untreated case reflects similar functional activity, achieved in the treated group by increasing epistasis that compensates for the information loss due to increased sequence variability. The increase in epistasis thus allows the protein to adapt and attain high fitness (via wild-type level biological activity), in the increasingly rugged fitness landscape of treatment.

Besides a trend in the overall strength of epistatic interactions in the treated group, we also find that the spatial organization of interactions in the protein is modified. In \nameref{sec:space}, we show that untreated subjects stored more information in distant pairs early (1998, distant: residue distance $\geq8\AA$), a trend that is reversed in 2002.  

%Fig. 3
%Fig. 4

As pairwise mutual information is a proxy for epistasis (Materials and Methods), the significant temporal increase in mutual information suggests that epistatic interactions are crucial for the protease to adapt in a dynamic drug environment. In contrast, the sum of pairwise mutual information remains fairly constant in the drug-free environment (Fig~\ref{fig:infoplot}, middle panel). It should be noted that most of the treated data for year 2003 came from two phase-III clinical trials that focused on the 2nd-line anti-retroviral drug tipranavir (2900 out of 3399 sequences)~\cite{Baxteretal2006}. Resistance to tipranavir requires accumulation of several mutations, more than the mutations required for the 1st-line protease inhibitors, and this higher genetic barrier to resistance makes it suitable for salvage therapy for patients already experiencing resistance to other drugs. The substantial decrease in $I_1$ for the year 2003 thus can be attributed to an increased entropy as a consequence of accumulating a higher number of mutations required for resistance to tipranavir. It is curious that the marked increase in variability in the tipranavir-dominated data set is not associated with a marked increase in pair-wise information (compared to the adjacent years), suggesting that the tipranavir-induced mutations are mostly non-epistatic. 
 
\begin{figure*}[htbp] % 
   \centering
   \includegraphics[width=5in]{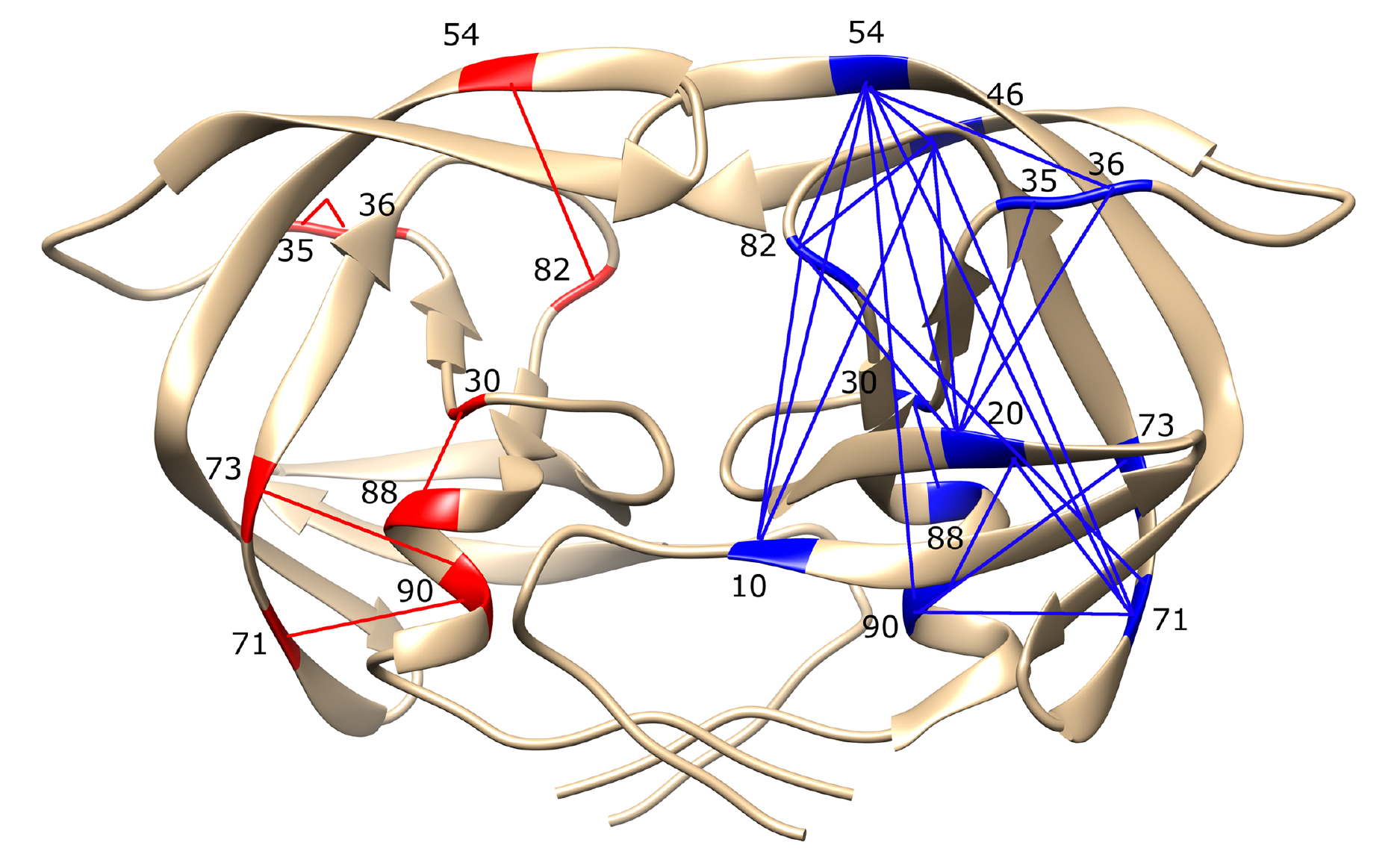} 
   \caption{{\bf Epistatic interactions mapped onto the protease structure.} Epistatic interactions in the protease sequences in treated data from the year 1998 (red on the left chain) and 2006 (blue, right chain). The interacting residues are numbered. Only those interactions are shown where information is greater than 0.1 bits, indicating strong epistasis. Figure generated using Chimera~\cite{Pettersenetal2004}.}
   \label{fig:protease}
\end{figure*}

Pairwise interactions between amino acids are thought to be sufficient to encode the protein fold, and thus pairwise interactions can be considered as a sufficient source of protein functional information~\cite{Socolichetal2005}. While there is currently no evidence that higher-order interactions between residues are important, some authors have discussed this issue~\cite{Weinreichetal2013}. Mapping the high information loci (interactions of a strength of at least 0.1 bits) on the protease structure for the treated group for the first year (1998) and comparing them to the treated group in 2006 (Fig~\ref{fig:protease}) clearly shows the increase in epistatic connections in the latter time point.  

%Fig. 5

The observation that strong selection due to treatment increases residue variability (thus decreasing $I_1$) while increasing epistatic interactions in the protease (as measured by pairwise mutual information) is corroborated by a longitudinal analysis of information measures in protease sequences from patients who were untreated at the first time point and received treatment at the second time point (see~\nameref{sec:long}). 

\section*{Discussion\label{sec:Discussion}}

It is difficult to predict the evolutionary trajectory of a protein from fitness effects of mutations along with mutation rate and population size, but we can trace the evolutionary history of a protein to understand the processes underlying its long-term evolution. While evolution experiments with bacteria, viruses, and yeast provide direct evidence for evolution-in-action, enabling the study of various aspects of evolutionary dynamics as well as the likelihood  of certain evolutionary paths~\cite{ElenaLenski2003, Barricketal2009, Poelwijketal2007}, indirect evidence such as sequence data collected over years can be a valuable resource to retrace the evolutionary steps taken by a protein on an adaptive fitness landscape over a long period of time. The sequences of the HIV-1 protease are one such resource that contain the ``fossils" of the evolutionary trajectory (albeit in a statistical form) of the protein. Although this sequence data comes from isolates collected from different patients over the years, the star-like phylogeny observed for data from each year indicates the absence of population subdivisions and linkage disequilibrium that might have confounded the information-theoretic analysis we present here (\nameref{sec:tree}). It is likely that each sequence in the data represents the major HIV-1 variant circulating in each patient. If the viral populations within the hosts have not yet reached equilibrium (if such an equilibrium exists in the presence of persistent selection pressures of host immune response and treatment), the sequence obtained from treated individuals will not contain all of the resistance and compensatory mutations that might be found at the equilibrium. Thus, the information-theoretic measures computed here represent a lower bound of the true sequence variability and epistatic interactions present in a within-host viral population.

We have investigated the long-term evolution of a protein by computational analysis of HIV-1 protease sequences from treated and untreated patients collected over a span of nine years (1998-2006), and find that the molecule responds to the selection pressures of treatment by accumulating mutations that confer drug resistance. At the same time, several positions in the protease show considerable neutrality even in the absence of treatment (Figure \ref{fig:1998_2006_entropies}).  Indeed, a complete mutagenesis of the protease showed that several sites are insensitive to mutation in the absence of a selection pressure, and thus appear neutral~\cite{Loebetal1989}. Yet, some of those neutral mutations often appear in tandem with known resistance mutations, possibly compounding the fitness effects of the resistance mutations~\cite{Velazquez-Campoyetal2003}. Even known resistance mutations usually do not confer resistance in isolation, but require compensatory mutations before resistance is achieved~\cite{Mammanoetal2000}. Because on average a mutation destabilizes the protein fold by about 1 kcal/mol, proteins cannot accumulate multiple resistance mutations without running the risk of thermal instability~\cite{Bloometal2005}. Yet, HIV-1 protease with resistance to multiple drugs can accumulate more than 10 mutations~\cite{Louisetal2013, Muzammiletal2003, Agniswamyetal2012}. We suspect that many of these mutations are {\em re-stabilizing} the protein,thus compensating for the fitness cost of resistance mutations~\cite{ChangTorbett2011,Arayaetal2012}. In addition, sequence logos of protease sequences from untreated patients (\nameref{sec:logo}) shows that several resistance and compensatory mutations are present at low frequencies prior to treatment, suggesting that pre-existing compensatory mutations may protect the protein from incurring the potentially strong fitness costs of resistance mutations. We show that as sequence variability increases due to treatment, more and more of the variable residues interact with other residues, leading to an increase in the epistatic interactions. Along this evolutionary trajectory, the protein finds itself in a fitness landscape that is increasingly rugged~\cite{Kouyosetal2012}.

That the ruggedness of the fitness landscape has significant effects on the evolution of a protein has been discussed extensively~\cite{KauffmanWeinberger1989,Frankeetal2011,KvitekSherlock2011,Ostmanetal2012,DraghiPlotkin2013,Nahumetal2015}. These studies suggest that epistasis can have either an inhibitory or an accelerating effect on evolutionary trajectories, determined mainly by whether evolution occurs far off the fitness peak or close to it. Computational simulations of evolutionary adaptation reveal that while increased epistasis correlates with high ruggedness in the fitness landscape, evolvability (the ability to attain the highest fitness peak) declines beyond a threshold epistasis~\cite{Ostmanetal2012}. The sign of epistasis (positive vs. negative epistasis) also can affect the speed of adaptation. Epistasis is positive (or negative) when the fitness effect of the double mutant is greater (or smaller) than the sum of the fitness effects of individual mutations. Negative epistasis between beneficial mutations is often associated with diminishing returns in climbing a single peak~\cite{Khanetal2011,Chouetal2011,Tokurikietal2012,Kryazhimskiyetal2014}, while positive epistasis is associated with compensatory effects~\cite{Wilkeetal2003} that can enhance crossing of valleys between peaks~\cite{Phillips2008,Ostmanetal2012}. Computational studies further clarify that positive epistasis accelerates evolvability, while negative epistasis promotes robustness~\cite{Carteretal2005}. However, if the landscape is too rugged, the reciprocal sign epistasis between mutations may prevent valley crossing~\cite{Chiottietal2014}. As we cannot determine the sign of epistasis using our information-theoretic methods, we are unable to verify that the epistasis between mutations in the HIV-1 protease is mostly positive, as suggested by Bonhoeffer et al.~\cite{Bonhoefferetal2004}.

We find here that as epistasis between residues in the protease continues to increase (as revealed by the monotonically increasing sum of pairwise mutual information), epistasis is still contributing to adaptation rather than inhibiting it. However, there is little doubt that evolution in a more and more rugged fitness landscape will have a significant impact on evolvability. For example, it is known that some compensatory mutations that repair fitness costs increase viral fitness in the absence of drugs~\cite{Theysetal2012}. These mutations are found in individuals who have transmitted drug resistance, but they are not common in the treatment-naive HIV-1 population. One suggestion is that the evolutionary pathway to this polymorphism is simply too complex to emerge de novo~\cite{Arts2012}, suggesting that the virus has evolved into a region of genetic space from which there is no return. 

Previous studies of epistatic interactions in adaptive evolution emphasized their importance for adaptation~\cite{Trindadeetal2009,GongBloom2014,Lagatoretal2014,Kryazhimskiyetal2014,Lozovskyetal2009,Ortlundetal2007,Karasovetal2010}. Our analysis provides statistical arguments about the changes in epistatic interactions during long-term adaptation of a protein in an environment of persistent high selective pressure. Ideally, the present findings should be corroborated by longitudinal studies that collect sequence or fitness data regularly and for an extended period of time. Such studies would significantly contribute to and understanding of the impact of strong selection pressures and continually changing landscapes on adaptive evolution.

\section*{Materials and Methods \label{sec:Methods}}
\subsection*{HIV-1 Protease sequences}
The protease sequences for HIV-1 subtype B were collected from the HIV Stanford database [http://hivdb.stanford.edu] on September 17, 2013. The database collects sequences from pilot studies and clinical trials that are published with sequence data deposited in GenBank. We only used sequences for which the exact dates for isolate collection are known (as opposed to guessed from the time of publication or clinical trials). 

We focused our analysis on 18,571 sequences from the years 1998 to 2006, in which more than 300 protease sequences are available for both treated and untreated sequence datasets (Tab~\ref{nseq}). Since these data come from different sources, they are not longitudinal. Sequences obtained from patients receiving $\ge$ 1 protease inhibitors are labeled as treated, while sequences from patients not receiving any protease inhibitors are labeled as untreated. Note that a fraction of new infections have transmitted drug-resistance, and thus even higher sequence diversity ($\approx$8\% of new cases had transmitted drug-resistance by January 2007~\cite{Rossetal2007}). Since the HIV Stanford database does not keep data on transmitted drug-resistance for treatment-naive individuals, we label all sequences from treatment-naive patients as `untreated' but cannot exclude the possibility that they might carry resistance mutations. However, due to their high fitness cost some drug-resistance mutations tend to revert back to wild-type residue in absence of therapy~\cite{Wangetal2011}, suggesting that the protease sequence diversity due to resistance mutations should decrease in cases of transmitted drug-resistance in treatment-naive patients. 
%Thus, an undetermined (but small) fraction of the data labelled `untreated' can contain some resistance mutations.

\begin{table}[!htbp]
\caption{{\bf Number of protease sequences in treated and untreated datasets.} Data downloaded from HIV Stanford database on September 17, 2013. The set of sequences we use for our analysis is available from Dryad.}
\label{nseq}
\center
\begin{tabular*}{0.5\textwidth}{@{\extracolsep{\fill}} |c c c |}%{@{\extracolsep{\fill}}llc{6,0}d{6,0}@{}}
\hline
Year & Untreated & Treated \\
\hline
1998& 376& 680 \\
1999& 1145& 946 \\
2000& 375& 806 \\
2001& 476& 575 \\
2002& 1982& 464 \\
2003& 1237& 3399 \\
2004& 2163& 436 \\
2005& 1424& 338 \\
2006& 1354& 341 \\
\hline
\end{tabular*}
\end{table}

We also obtained longitudinal data (protease sequences collected from the same patient after a time interval) that was available from HIV Stanford database for the following: i) patient was untreated at first as well as second isolate collection (untreated to untreated): 596 sequences, ii) patient untreated at first isolate collection but treated with protease inhibitors at second isolate collection (untreated to treated): 395 sequences, iii) patient treated at both isolate collections (treated to treated): 921 sequences, and iv) patient treated at first isolate collection but untreated at second isolate collection (treated to untreated): 153 sequences. The time between first and second isolate collections ranged from one month to a few years.

\subsection*{Information as a proxy for epistasis}
Positive information between two sites indicates that the two sites are co-varying, but co-variation is not the same as epistasis. To study the relationship between informational co-variation and epistasis, we constructed a population-genetic two alleles-two loci model that we solve using replicator-mutator equations. (The model can be generalized to two loci with 20 alleles in a straightforward manner.)

\begin{figure}[htbp] %  
   \centering
   \includegraphics[width=3in]{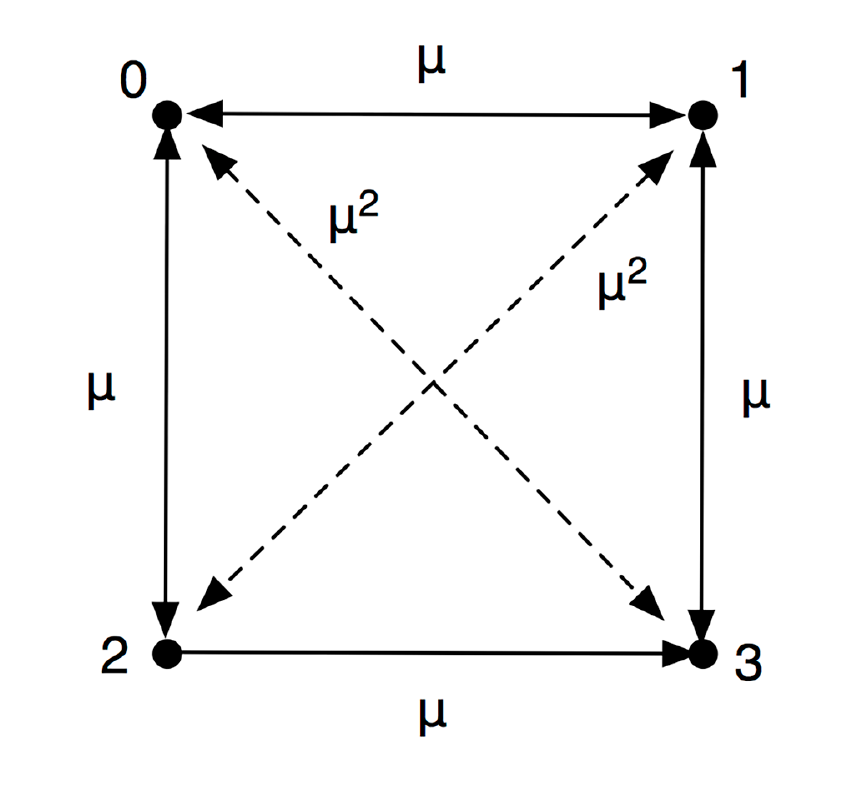} 
   \caption{{\bf Rates of mutation between four different genotypes}. The types are denoted as 0=AA, 1=Aa, 2=aA, and 3=aa.}
   \label{Sfig1}
\end{figure}

We consider four genotypes (two alleles `A' and `a', at two loci): AA: (type 0, the wild type),  Aa (type 1), aA (type 2), and aa: (type 3) that undergo mutation with rate $\mu$ per unit time (see Fig~\ref{Sfig1}).
%Fig. 6
The probability to find each of these genotypes in an infinite population depends on the fitness and probabilities of the other genotypes. In a discrete update scheme, the probability to find type $i$ at time $t+1$ is related to the same quantity at time $t$ via
\be
p_0^{t+1}&=&p_0^t\frac{w_0}{\bar w} F +\mu\left(\frac{p_1^tw_1+p_2^tw_2}{\bar w}\right)+\mu^2 \frac{p_3^tw_3}{\bar w}  \label{req1}\\
p_1^{t+1}&=&p_1^t\frac{w_1}{\bar w} F +\mu\left(\frac{p_0^tw_0+p_3^tw_3}{\bar w}\right)+\mu^2 \frac{p_2^tw_2}{\bar w} \label{req2}\\
p_2^{t+1}&=&p_2^t\frac{w_2}{\bar w} F +\mu\left(\frac{p_0^tw_0+p_3^tw_3}{\bar w}\right)+\mu^2 \frac{p_1^tw_1}{\bar w} \label{req3}\\
p_3^{t+1}&=&p_3^t\frac{w_3}{\bar w} F +\mu\left(\frac{p_1^tw_1+p_2^tw_2}{\bar w}\right)+\mu^2 \frac{p_0^tw_0}{\bar w} \label{req4}
\ee
where $\bar w$ is the mean fitness $\bar w = \sum_{i=0}^3 p^t_i w_i$, and $F$ is the fidelity of replication $F=1-2\mu-\mu^2$. 
It is easy to show that $\sum p_i^{t+1}=1$ as long as $\sum p_i^{t}=1$.

\begin{figure}[htbp] %  
   \centering
  \includegraphics[width=4in]{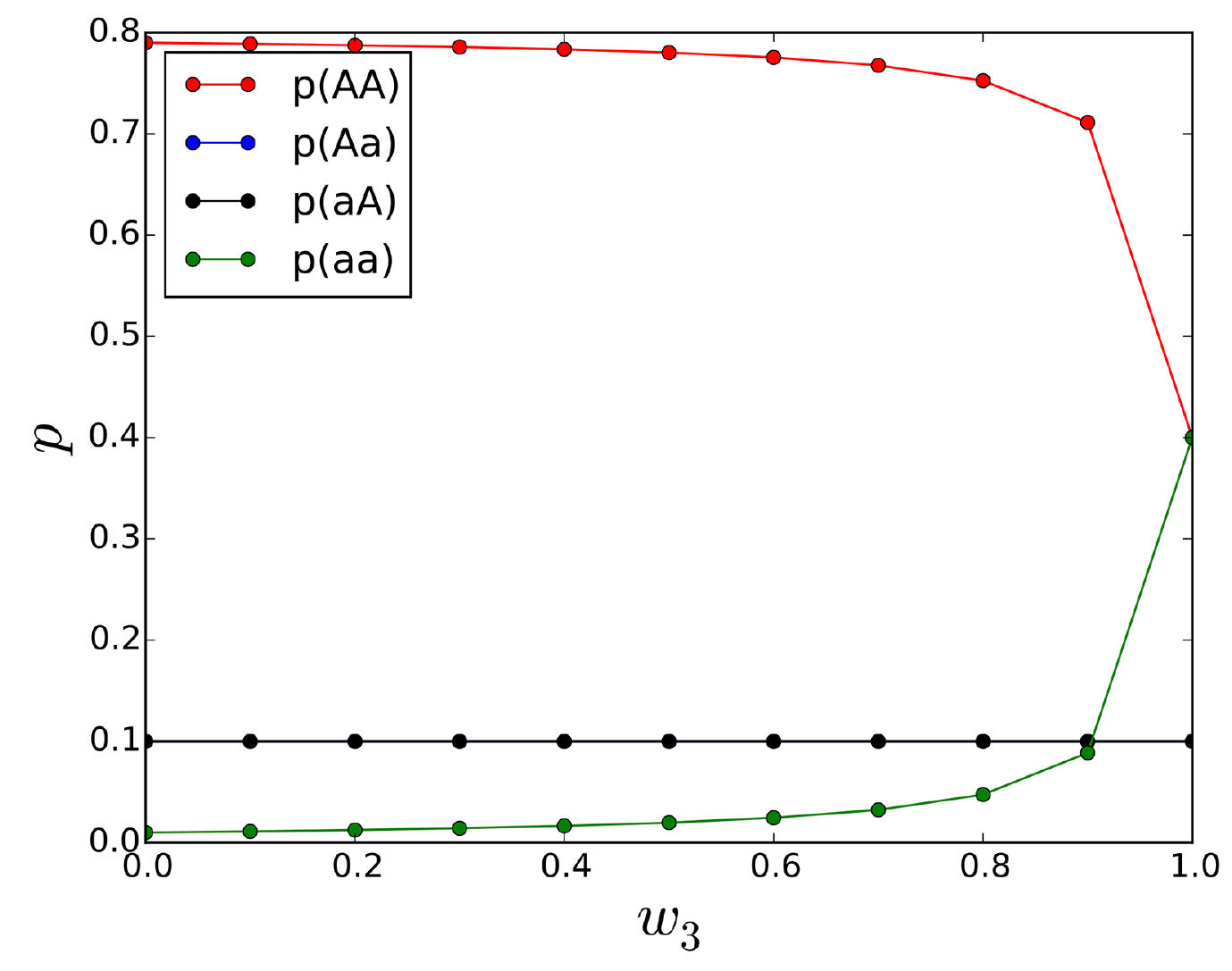} 
   \caption{{\bf Allele frequencies as fitness for type 3 (aa) is varied}. In this valley-crossing landscape, $w_0$ is always 1 and $w_1$ = $w_2$ = 0. Plot shows allele frequencies $p_i$ at mutation rate $\mu=0.1$ as a function of $w_3$. The intermediate types aA and Aa occur only at the rate of mutation as they have zero fitness.}
   \label{genprob2}
\end{figure}

Equations (\ref{req1}-\ref{req4}) can be solved numerically iteratively, but alternatively the fixed point (the $p_i$ in the limit $t\to\infty$) can be calculated by solving for the right eigenvector of the associated Markov matrix. We investigate different fitness landscapes by varying the fitness of the mutants, while the fitness of the wild-type is constant at $w_0=1$. Fig~\ref{genprob2} shows equilibrium allele frequencies for a landscape where the double mutant has a given fitness $w_3$, while the intermediate genotypes have zero fitness (a valley-crossing landscape with reciprocal sign epistasis).  

%Fig. 7

Armed with the equilibrium probabilities $p_i$, we can calculate the information between loci as follows. First we define $p(A)$ and $p(a)$ for the first and second locus:
\be p^{(1)}(A)=p_0+p_1\;,\;\;\; p^{(1)}(a)=1-p_0-p_1\nonumber \\
p^{(2)}(A)=p_0+p_2\;,\;\;\; p^{(2)}(a)=1-p_0-p_2
\ee
giving us the marginal entropies of the first and second locus
\be
H(1)&=&-\sum_{i=a,A}p^{(1)}(i)\log p^{(1)}(i)\;, \nonumber \\
H(2)&=&-\sum_{i=a,A}p^{(2)}(i)\log p^{(2)}(i)\;.
\ee
The joint entropy of both loci is then
\be
H(1,2)=-\sum_{i=a,A}\sum_{j=a,A}p^{(1,2)}(i,j)\log p^{(1,2)}(i,j)\;.
\ee
where $p^{(1,2)}(i,j)$ is the joint probability to observe allele $i$ at locus 1 and allele $j$ at locus 2. \\
The shared entropy (or information)  is
\be
I(1:2)=H(1)+H(2)-H(1,2)\;.
\ee
We can then relate this information to the epistasis between loci calculated as~\cite{Bonhoefferetal2004,Ostmanetal2012}
\be
E=\log(\frac{w_3}{w_0})-\log(\frac{w_2}{w_0})-\log(\frac{w_1}{w_0})=\log(\frac{w_3 w_0}{w_1 w_2}) \label{epieqn}
\ee
There are other ways of defining epistasis between loci (see, e.g.,~\cite{Phillips2008}), 
but the qualitative relation between information and epistasis is not affected. 

\begin{figure}[htbp] %  
   \centering  \includegraphics[width=5in]{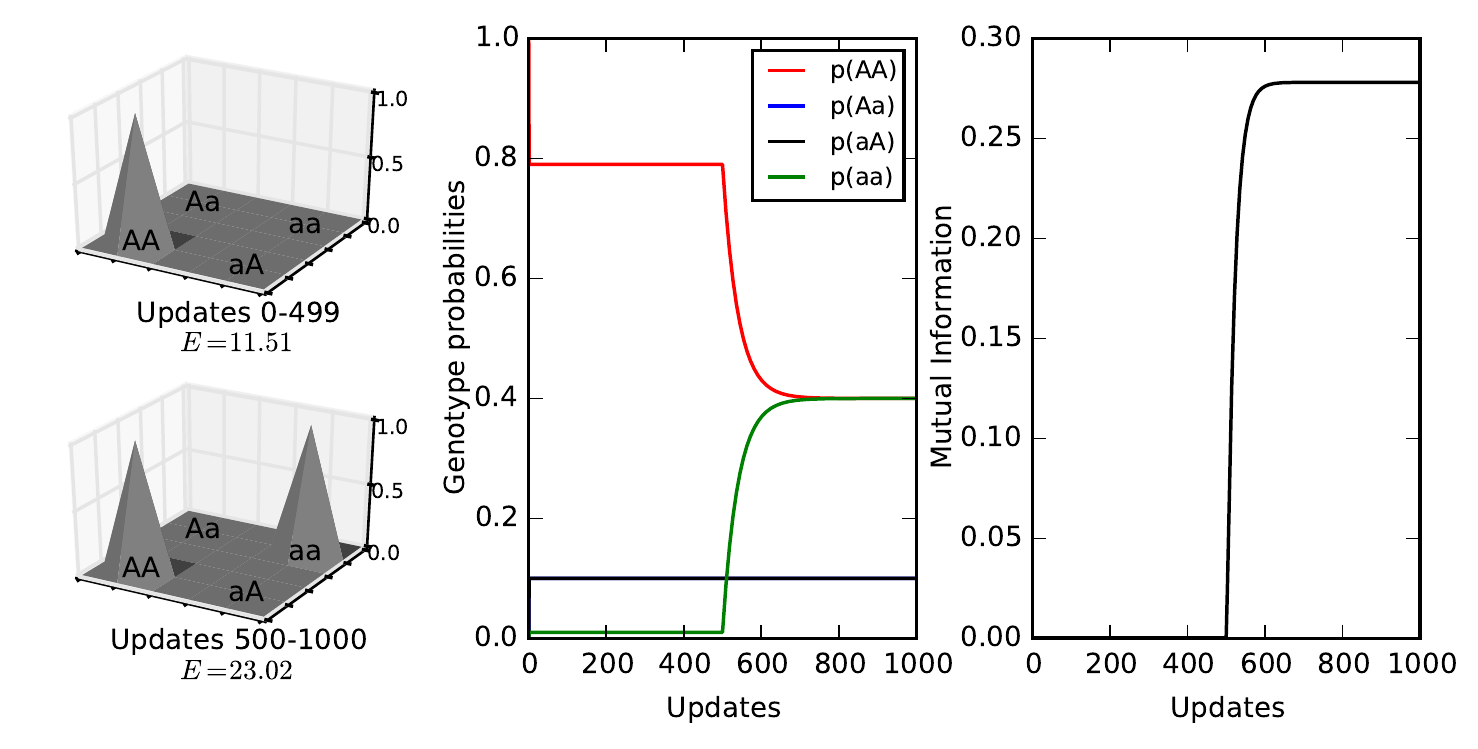} 
   \caption{{\bf Two-loci two-allele model}. The left panel shows the fitness landscapes and epistasis given by Eq.~(\ref{epieqn}) in the first and second half of the simulation (updates 0-499:  $w_0=1$ and $w_1=w_2=w_3=10^{-5}\approx0$; updates 500-1000: $w_0=w_3=1$ and $w_1=w_2=10^{-5}\approx0$). The xy-plane shows the four genotypes while the z-axis shows genotype fitness. The middle panel shows the genotype probabilities while the right panel shows the mutual information during the course of the simulation. Note that the increase in epistasis at the 500th update is reflected in the increase in mutual information. The mutation rate was 0.1 and starting population frequencies were $p_0=1$ and $p_1=p_2=p_3=0$.}
   \label{fig-twoloci}
\end{figure}

An extreme example occurs when $w_0=w_3=1$ and $w_1=w_2=0$, that is, when the double mutant has the same fitness as the wild type, but the intermediate genotypes have no fitness. In that case, it is necessary to cross a valley in the fitness landscape to reach the double mutant $aa$. In this case of reciprocal sign epistasis~\cite{Poelwijketal2007}, $E=\infty$, and 
\begin{equation}
I(1:2)=-(1-p_0)\log(1-p_0)-(1-p_3)\log(1-p_3)\;.
\end{equation}If $p_0=p_3=0.5$ (full equilibration) this extreme level of epistasis correspond to 1 bit of information (the maximum possible). Fig.~\ref{fig-twoloci} shows the changes in genotype probabilities and mutual information as the population adapts from a single-peak landscape ($w_0=1$ and $w_1=w_2=w_3\approx0$) to a two-peak fitness landscape landscape ($w_0=w_3=1$ and $w_1=w_2\approx0$). Pairwise mutual information increases as the landscape becomes more rugged. See \nameref{sec:threeloci} for simulations for a three-loci two-allele model that show that an increase in the sum of mutual information coincides with the increase in the ruggedness of the fitness landscape.

%Fig. 8

\begin{figure}[htbp] %  
   \centering
  \includegraphics[width=4in]{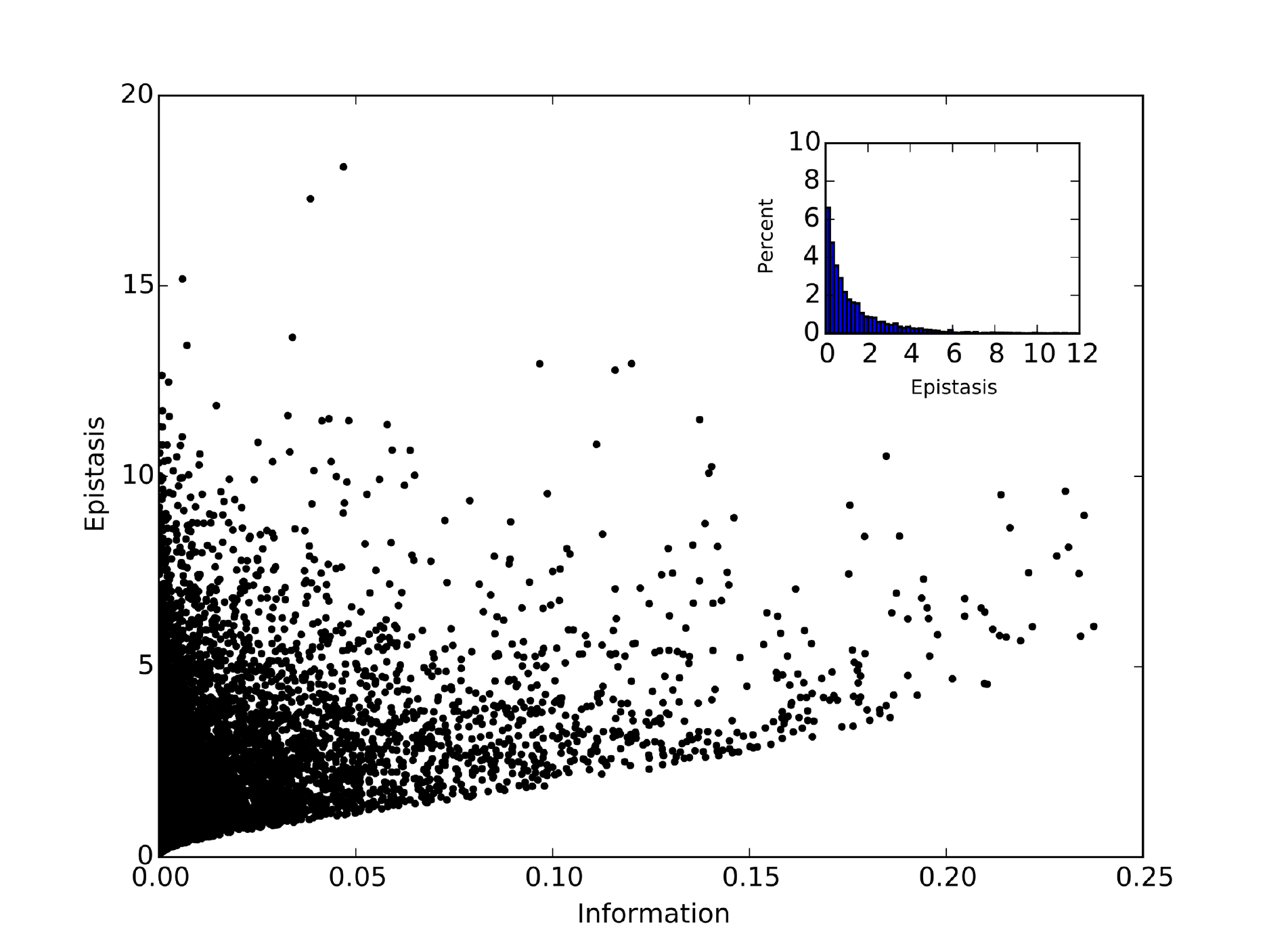} 
   \caption{{\bf Correlation between epistasis and information}. Each point corresponds to information and absolute value of epistasis calculated for one of the 10,000 combinations of $w_0$, $w_1$, $w_2$, and $w_3$. Here, $w_0$ is always 1, and other fitness values are uniformly randomly assigned between 0 and 1. The inset shows the percentage of points with negligible information (\textless 0.001 bits) as a function of epistasis.}
   \label{fig-spear}
\end{figure}

To study the general relationship between epistasis and information, we calculated both epistasis and information starting with $w_0=1$ (wild type fitness), and random fitness values (between 0 and 1) for the single and double mutants $w_1$, $w_2$, and $w_3$. The equilibrium genotype probabilities were obtained by iterating the replicator-mutator equations until they had stabilized (30,000 updates of genotype probabilities $p_0$, $p_1$, $p_2$, and $p_3$ using equations (\ref{req1}-\ref{req4}), with starting genotype probabilities: $p_0 = 1$, $p_1 = p_2 = p_3 = 0$). We find that the absolute value of epistasis $|E|$ is positively correlated with information (Spearman $R=0.497$, $p<10^{-15}$, see Fig~\ref{fig-spear}). 
%Fig. 9
It is clear that information is a sufficient (but not necessary) condition for epistasis. Thus, high information between two loci guarantees epistasis between those two loci, but there may be epistatically interacting loci that are missed when focusing only on information, as some loci can interact epistatically without being informative about each other. In addition, the direction of epistasis (the sign of $E$) cannot be determined solely from information. To ascertain the fraction of epistatically interacting pairs that are missed by an informational analysis, we examined pairs with information below a chosen cut-off (here $I<0.001$ bits). About 35\% of all pairs have information below the threshold, but only 28\% of all pairs have non-vanishing epistasis at the same time (see inset in Fig~\ref{fig-spear}, which shows the distribution of epistasis among pairs with sub-threshold information). This finding leads us to conclude that, at least in the simple model presented here, only about 30\% of epistatically interacting pairs are missed by using an information proxy. Furthermore, the majority of those missed pairs have small epistasis, as evidenced by the distribution of epistasis for those pairs with negligible information in the inset of Fig~\ref{fig-spear}.

\subsection*{Information stored in HIV-1 protease}
The information content of biomolecules can be estimated using information-theoretic constructs~\cite{AdamiCerf2000,Adami2004,Adami2012}. Information content is different from sequence length: it can be thought of as the amount of information that is stored in the protein sequence about the cellular environment within which it functions (implying that it is contextual).
Information content can change when the environment changes even if the sequence remains the same, and is thought to be a proxy for fitness~\cite{RivoireLeibler2011}. Because a changed environment usually translates into reduced information content (reflecting reduced fitness), the virus seeks to recover the information and achieve previous fitness levels by evolving drug resistance. 

The total information content of a protein of length $\ell$ (taking into account all correlations between residues) is given by (adapting a formula for the ``multi-information" of $\ell$ events due to Fano~\cite{Fano1961}):
\begin{equation}
I_\ell = H_{\rm max} - \left(  \sum_{i=1}^\ell H_i - \sum_{i < j}^\ell I(i : j) + \sum_{i < j < k}^\ell I(i : j : k) - \dots \right)\;, \label{full}
\end{equation}
where $I(i:j)$ is the mutual information between two sites, $I(i:j:k)$ the information shared between three sites~\cite[p.~80] {Fano1961}, and so on. The terms not shown in Eq.~(\ref{full}) are the higher-order corrections $I(i:j:k:m)$ etc., up to $I(i_1:i_2:...:i_\ell)$, with alternating signs. Corrections due to interactions between three or more sites are expected to be small, but cannot be estimated using the present data because the samples are too small. $H_{\rm max}$ is the sum of maximum entropy at every site, and thus is equal to $\log_2(20)\times\ell\approx4.32\times \ell$ bits, where $\ell$ is the sequence length~\cite{Adami2004}. 

We define the first- and second-order information estimates of (\ref{full}) $I_1$ and $I_2$ as follows:
\begin{equation}
I_1 = H_{\rm max} - \sum_{i=1}^l H_i \label{i1}\;,
\end{equation} 

\begin{equation}
I_2 = I_1+ \sum_{i < j}^\ell I(i : j) \label{pairwise}\;.
\end{equation} 
Thus, $I_1$ measures the information content of the protein without considering any interactions among protein residues, and $I_2$ includes information contained in pairwise interactions between protein positions over and above $I_1$, but ignores any other higher-order interactions between residues. The second term in Eq.~(\ref{pairwise}) is the sum of pairwise mutual information for all pairs of residues in the protein. Note that if information was only described by $I_1$, mutations that provide resistance {\em reduce} the information in the ensemble, as they increase sequence entropy. Achieving an increase in total information must then occur via correlated mutations. 

\subsection*{Changes in physicochemical properties of the residues}
The site-wise changes in physicochemical properties of residues such as the iso-electric point pI and residue weight in the protease are calculated by averaging their values across the subsamples from each year and environment (treated and untreated).  The physicochemical properties of residues and their changes are discussed in~\nameref{sec:chem}.

\subsection*{Finite-size corrections for entropy estimates}
Small datasets do not correctly estimate the amino acid probabilities due to dataset-dependent observed frequencies of residues: this introduces a bias in the entropy and mutual information calculations (see, e.g.,~\cite{Basharin1959,Nemenmanetal2002}). Several priors and estimators have been proposed to estimate entropies from undersampled probability distributions~\cite{Nemenmanetal2002}, and our analysis suggests that a sample size of 300 with NSB (Nemenman-Shafee-Bialek) entropy bias correction gives reliable estimates of entropy and mutual information values (see \nameref{sec:bias}). 

Since the number of treated and untreated protease sequences in our dataset is different across years (Tab~\ref{nseq}), we sampled (with replacement) 10 sets/year of 300 sequences each to account for sample size bias. We calculate bias-corrected entropies and pairwise mutual information for the subsampled datasets, and report the average values (along with $\pm 1$ SD). Because several protease sequences had gaps at the sequence ends, we calculated the $I_1$, $I_2$, and sum of pairwise mutual information for a truncated sequence length (from positions 15 to 90 instead of positions 1 to 99 of the protease sequence, see~\nameref{sec:numbers}).

\subsection*{Statistical analysis of temporal trend of protein information}
To compare the temporal trends of $I_1$, $I_2$, and sum of pairwise mutual information for untreated and treated sequence data, we first fit linear regression models to the yearly data using the R statistical analysis platform~\cite{R-Core-Team2013} and then determine if the slopes of the regression models (for treated and untreated datasets) are significantly different or not. 

Although the data used in this analysis is not longitudinal (protease sequences are collected from different patients participating in different clinical trials/studies across the years), the linear regression model is fit to the average values of the response variables ($I_1$, $I_2$, and sum of pairwise mutual information) calculated by sampling ten sets of 300 sequences each, and thus reflect the approximate temporal trends of the response variables with respect to the two factors (untreated and treated).

\subsection*{Data Availability}
The sequence data used, analysis scripts, and the iPython notebook for generating figures are available in the Dryad repository: http://doi.org/10.5061/dryad.q66s5.

\section*{Acknowledgments}
%This work was supported by the National Science Foundation' BEACON Center for the Study of Evolution in Action, under contract No. DBI-0939454. 
We thank Jo\~ao Martins for collaboration in the early stages of this project. We wish to acknowledge the support of the Michigan State University High Performance Computing Center and the Institute for Cyber Enabled Research. 

%\bibliography{protease_paper_plos}

\begin{thebibliography}{10}

\bibitem{deVisseretal2011}
de~Visser JAGM, Cooper TF, Elena SF.
\newblock The causes of epistasis.
\newblock Proc Roy Soc B. 2011;278:3617--24.

\bibitem{Ortlundetal2007}
Ortlund EA, Bridgham JT, Redinbo MR, Thornton JW.
\newblock Crystal structure of an ancient protein: evolution by conformational
  epistasis.
\newblock Science. 2007;317:1544--8.

\bibitem{Tokurikietal2008}
Tokuriki N, Stricher F, Serrano L, Tawfik DS.
\newblock How protein stability and new functions trade off.
\newblock PLoS Comput Biol. 2008;4:e1000002.

\bibitem{SoskineTawfik2010}
Soskine M, Tawfik DS.
\newblock Mutational effects and the evolution of new protein functions.
\newblock Nat Rev Genet. 2010;11:572--82.

\bibitem{Bershteinetal2006}
Bershtein S, Segal M, Bekerman R, Tokuriki N, Tawfik DS.
\newblock Robustness-epistasis link shapes the fitness landscape of a randomly
  drifting protein.
\newblock Nature. 2006;444:929--32.

\bibitem{Banketal2015}
Bank C, Hietpas RT, Jensen JD, Bolon DNA.
\newblock A systematic survey of an intragenic epistatic landscape.
\newblock Mol Biol Evol. 2015;32:229--38.

\bibitem{HaydenWagner2012}
Hayden EJ, Wagner A.
\newblock Environmental change exposes beneficial epistatic interactions in a
  catalytic RNA.
\newblock Proc Roy Soc B. 2012;279:3418--25.

\bibitem{Maisnier-PatinAndersson2004}
Maisnier-Patin S, Andersson DI.
\newblock Adaptation to the deleterious effects of antimicrobial drug
  resistance mutations by compensatory evolution.
\newblock Res Microbiol. 2004;155:360--9.

\bibitem{Flynnetal2013}
Flynn KM, Cooper TF, Moore FBG, Cooper VS.
\newblock The environment affects epistatic interactions to alter the topology
  of an empirical fitness landscape.
\newblock PLoS Genet. 2013;9:e1003426.

\bibitem{MartinezPicado2008}
Martinez-Picado J, Mart{\'\i}nez MA.
\newblock HIV-1 reverse transcriptase inhibitor resistance mutations and
  fitness: a view from the clinic and ex vivo.
\newblock Virus Res. 2008;134:104--23.

\bibitem{Bloometal2005}
Bloom JD, Silberg JJ, Wilke CO, Drummond DA, Adami C, Arnold FH.
\newblock Thermodynamic prediction of protein neutrality.
\newblock Proc Natl Acad Sci USA. 2005;102:606--611.

\bibitem{ChangTorbett2011}
Chang MW, Torbett BE.
\newblock Accessory mutations maintain stability in drug-resistant {HIV-1}
  protease.
\newblock J Mol Biol. 2011;410:756--760.

\bibitem{Arayaetal2012}
Araya CL, Fowler DM, Chen W, Muniez I, Kelly JW, Fields S.
\newblock A fundamental protein property, thermodynamic stability, revealed
  solely from large-scale measurements of protein function.
\newblock Proc Natl Acad Sci U S A. 2012;109:16858--63.

\bibitem{Trindadeetal2009}
Trindade S, Sousa A, Xavier KB, Dionisio F, Ferreira MG, Gordo I.
\newblock Positive epistasis drives the acquisition of multidrug resistance.
\newblock PLoS Genet. 2009;5:e1000578.

\bibitem{HarmsThornton2014}
Harms MJ, Thornton JW.
\newblock Historical contingency and its biophysical basis in glucocorticoid
  receptor evolution.
\newblock Nature. 2014;512:203--7.

\bibitem{Wellneretal2013}
Wellner A, Raitses~Gurevich M, Tawfik DS.
\newblock Mechanisms of protein sequence divergence and incompatibility.
\newblock PLoS Genet. 2013;9:e1003665.

\bibitem{Lunzeretal2010}
Lunzer M, Golding GB, Dean AM.
\newblock Pervasive cryptic epistasis in molecular evolution.
\newblock PLoS Genet. 2010;6:e1001162.

\bibitem{Lozovskyetal2009}
Lozovsky ER, Chookajorn T, Brown KM, Imwong M, Shaw PJ, Kamchonwongpaisan S,
  et~al.
\newblock Stepwise acquisition of pyrimethamine resistance in the malaria
  parasite.
\newblock Proc Natl Acad Sci U S A. 2009;106:12025--30.

\bibitem{Borrelletal2013}
Borrell S, Teo Y, Giardina F, Streicher EM, Klopper M, Feldmann J, et~al.
\newblock Epistasis between antibiotic resistance mutations drives the
  evolution of extensively drug-resistant tuberculosis.
\newblock Evol Med Public Health. 2013;2013:65--74.

\bibitem{Lagatoretal2014}
Lagator M, Colegrave N, Neve P.
\newblock Selection history and epistatic interactions impact dynamics of
  adaptation to novel environmental stresses.
\newblock Proc Roy Soc B. 2014;281:20141679.

\bibitem{GongBloom2014}
Gong LI, Bloom JD.
\newblock Epistatically interacting substitutions are enriched during adaptive
  protein evolution.
\newblock PLoS Genet. 2014;10:e1004328.

\bibitem{Kouyosetal2007}
Kouyos RD, Silander OK, Bonhoeffer S.
\newblock Epistasis between deleterious mutations and the evolution of
  recombination.
\newblock Trends Ecol Evol. 2007;22:308--15.

\bibitem{Kouyosetal2012}
Kouyos RD, Leventhal GE, Hinkley T, Haddad M, Whitcomb JM, Petropoulos CJ,
  et~al.
\newblock Exploring the complexity of the {HIV-1} fitness landscape.
\newblock PLoS Genet. 2012;8:e1002551.

\bibitem{Hinkleyetal2011}
Hinkley T, Martins J, Chappey C, Haddad M, Stawiski E, Whitcomb JM, et~al.
\newblock A systems analysis of mutational effects in {HIV-1} protease and
  reverse transcriptase.
\newblock Nat Genet. 2011;43:487--9.

\bibitem{Ostmanetal2012}
Ostman B, Hintze A, Adami C.
\newblock Impact of epistasis and pleiotropy on evolutionary adaptation.
\newblock Proc Royal Soc B. 2012;279:247--56.

\bibitem{Kryazhimskiyetal2014}
Kryazhimskiy S, Rice DP, Jerison ER, Desai MM.
\newblock Global epistasis makes adaptation predictable despite sequence-level
  stochasticity.
\newblock Science. 2014;344:1519--22.

\bibitem{Szameczetal2014}
Szamecz B, Boross G, Kalapis D, Kov{\'a}cs K, Fekete G, Farkas Z, et~al.
\newblock The genomic landscape of compensatory evolution.
\newblock PLoS Biol. 2014;12:e1001935.

\bibitem{Quandtetal2014}
Quandt EM, Deatherage DE, Ellington AD, Georgiou G, Barrick JE.
\newblock Recursive genomewide recombination and sequencing reveals a key
  refinement step in the evolution of a metabolic innovation in {{\em
  Escherichia coli}}.
\newblock Proc Natl Acad Sci U S A. 2014;111:2217--22.

\bibitem{BrikWong2003}
Brik A, Wong CH.
\newblock {HIV-1} protease: {M}echanism and drug discovery.
\newblock Org Biomol Chem. 2003;1:5--14.

\bibitem{ManskyTemin1995}
Mansky LM, Temin HM.
\newblock Lower in vivo mutation rate of human immunodeficiency virus type 1
  than that predicted from the fidelity of purified reverse transcriptase.
\newblock J Virol. 1995;69:5087--94.

\bibitem{Brownetal1999}
Brown AJ, Korber BT, Condra JH.
\newblock Associations between amino acids in the evolution of HIV type 1
  protease sequences under indinavir therapy.
\newblock AIDS Res Hum Retroviruses. 1999;15:247--253.

\bibitem{Hoffmanetal2003}
Hoffman NG, Schiffer CA, Swanstrom R.
\newblock Covariation of amino acid positions in {HIV-1} protease.
\newblock Virology. 2003;314:536--548.

\bibitem{Haqetal2009}
Haq O, Levy RM, Morozov AV, Andrec M.
\newblock Pairwise and higher-order correlations among drug-resistance
  mutations in HIV-1 subtype B protease.
\newblock BMC Bioinformatics. 2009;10 Suppl 8:S10.

\bibitem{Coffin1995}
Coffin JM.
\newblock {HIV} population dynamics in vivo: implications for genetic
  variation, pathogenesis, and therapy.
\newblock Science. 1995;267:483--9.

\bibitem{Pennings2012}
Pennings PS.
\newblock Standing genetic variation and the evolution of drug resistance in
  HIV.
\newblock PLoS Comput Biol. 2012;8:e1002527.

\bibitem{Mammanoetal1998}
Mammano F, Petit C, Clavel F.
\newblock Resistance-associated loss of viral fitness in human immunodeficiency
  virus type 1: Phenotypic analysis of protease and gag coevolution in protease
  inhibitor-treated patients.
\newblock J Virol. 1998;72:7632--7637.

\bibitem{Penningsetal2014}
Pennings PS, Kryazhimskiy S, Wakeley J.
\newblock Loss and recovery of genetic diversity in adapting populations of
  HIV.
\newblock PLoS Genet. 2014;10:e1004000.

\bibitem{Morand-Joubertetal2006}
Morand-Joubert L, Charpentier C, Poizat G, Ch{\^e}ne G, Dam E, Raguin G, et~al.
\newblock Low genetic barrier to large increases in HIV-1 cross-resistance to
  protease inhibitors during salvage therapy.
\newblock Antivir Ther. 2006;11:143--54.

\bibitem{Theysetal2012}
Theys K, Deforche K, Vercauteren J, Libin P, van~de Vijver DA, Albert J, et~al.
\newblock Treatment-associated polymorphisms in protease are significantly
  associated with higher viral load and lower CD4 count in newly diagnosed
  drug-naive HIV-1 infected patients.
\newblock Retrovirology. 2012;9:81.

\bibitem{TavernaGoldstein2002}
Taverna DM, Goldstein RA.
\newblock Why are proteins marginally stable?
\newblock Proteins. 2002;46:105--9.

\bibitem{BloomArnold2009}
Bloom JD, Arnold FH.
\newblock In the light of directed evolution: pathways of adaptive protein
  evolution.
\newblock Proc Natl Acad Sci U S A. 2009;106 Suppl 1:9995--10000.

\bibitem{BurchChao1999}
Burch CL, Chao L.
\newblock Evolution by small steps and rugged landscapes in the {RNA} virus
  $\phi$6.
\newblock Genetics. 1999;151:921--7.

\bibitem{KvitekSherlock2011}
Kvitek DJ, Sherlock G.
\newblock Reciprocal sign epistasis between frequently experimentally evolved
  adaptive mutations causes a rugged fitness landscape.
\newblock PLoS Genet. 2011;7:e1002056.

\bibitem{Adami2004}
Adami C.
\newblock Information theory in molecular biology.
\newblock Phys Life Rev. 2004;1:3--22.

\bibitem{Strelioffetal2010}
Strelioff CC, Lenski RE, Ofria C.
\newblock Evolutionary dynamics, epistatic interactions, and biological
  information.
\newblock J Theor Biol. 2010;266:584--94.

\bibitem{Anastassiou2007}
Anastassiou D.
\newblock Computational analysis of the synergy among multiple interacting
  genes.
\newblock Mol Syst Biol. 2007;3:83.

\bibitem{daSilva2009}
da~Silva J.
\newblock Amino acid covariation in a functionally important human
  immunodeficiency virus type 1 protein region is associated with population
  subdivision.
\newblock Genetics. 2009;182:265--75.

\bibitem{WangLee2007}
Wang Q, Lee C.
\newblock Distinguishing functional amino acid covariation from background
  linkage disequilibrium in HIV protease and reverse transcriptase.
\newblock PLoS One. 2007;2:e814.

\bibitem{Shaferetal2007}
Shafer RW, Rhee SY, Pillay D, Miller V, Sandstrom P, Schapiro JM, et~al.
\newblock HIV-1 protease and reverse transcriptase mutations for drug
  resistance surveillance.
\newblock AIDS. 2007;21:215--23.

\bibitem{Jakobsenetal2010}
Jakobsen MR, Tolstrup M, S{\o}gaard OS, J{\o}rgensen LB, Gorry PR, Laursen A,
  et~al.
\newblock Transmission of {HIV-1} drug-resistant variants: Prevalence and
  effect on treatment outcome.
\newblock Clin Infect Dis. 2010;50:566--573.

\bibitem{Yerlyetal1999}
Yerly S, Kaiser L, Race E, Bru JP, Clavel F, Perrin L.
\newblock Transmission of antiretroviral-drug-resistant {HIV-1} variants.
\newblock Lancet. 1999;354:729--733.

\bibitem{ShaferSchapiro2008}
Shafer RW, Schapiro JM.
\newblock HIV-1 drug resistance mutations: An updated framework for the second
  decade of HAART.
\newblock AIDS Rev. 2008;10:67--84.

\bibitem{Mammanoetal2000}
Mammano F, Trouplin V, Zennou V, Clavel F.
\newblock Retracing the evolutionary pathways of human immunodeficiency virus
  type 1 resistance to protease inhibitors: {V}irus fitness in the absence and
  in the presence of drug.
\newblock J Virol. 2000;74:8524--8531.

\bibitem{Sokalingametal2012}
Sokalingam S, Raghunathan G, Soundrarajan N, Lee SG.
\newblock A study on the effect of surface lysine to arginine mutagenesis on
  protein stability and structure using green fluorescent protein.
\newblock PLoS One. 2012;7:e40410.

\bibitem{Martinez-Picadoetal1999}
Martinez-Picado J, Savara AV, Sutton L, D'Aquila RT.
\newblock Replicative fitness of protease inhibitor-resistant mutants of human
  immunodeficiency virus type 1.
\newblock J Virol. 1999;73:3744--52.

\bibitem{Carothersetal2004}
Carothers JM, Oestreich SC, Davis JH, Szostak JW.
\newblock Informational complexity and functional activity of {RNA} structures.
\newblock J American Chem Society. 2004;126:5130--5137.

\bibitem{Baxteretal2006}
Baxter JD, Schapiro JM, Boucher CAB, Kohlbrenner VM, Hall DB, Scherer JR,
  et~al.
\newblock Genotypic changes in human immunodeficiency virus type 1 protease
  associated with reduced susceptibility and virologic response to the protease
  inhibitor tipranavir.
\newblock J Virol. 2006;80:10794--10801.

\bibitem{Socolichetal2005}
Socolich M, Lockless SW, Russ WP, Lee H, Gardner KH, Ranganathan R.
\newblock Evolutionary information for specifying a protein fold.
\newblock Nature. 2005;437:512--8.

\bibitem{Weinreichetal2013}
Weinreich DM, Lan Y, Wylie CS, Heckendorn RB.
\newblock Should evolutionary geneticists worry about higher-order epistasis?
\newblock Curr Opin Genet Dev. 2013;23:700--707.

\bibitem{Pettersenetal2004}
Pettersen EF, Goddard TD, Huang CC, Couch GS, Greenblatt DM, Meng EC, et~al.
\newblock {UCSF Chimera}--a visualization system for exploratory research and
  analysis.
\newblock J Comput Chem. 2004;25:1605--12.

\bibitem{ElenaLenski2003}
Elena SF, Lenski RE.
\newblock Evolution experiments with microorganisms: The dynamics and genetic
  bases of adaptation.
\newblock Nat Rev Genet. 2003;4:457--69.

\bibitem{Barricketal2009}
Barrick JE, Yu DS, Yoon SH, Jeong H, Oh TK, Schneider D, et~al.
\newblock Genome evolution and adaptation in a long-term experiment with {\it
  Escherichia coli}.
\newblock Nature. 2009;461:1243--7.

\bibitem{Poelwijketal2007}
Poelwijk FJ, Kiviet DJ, Weinreich DM, Tans SJ.
\newblock Empirical fitness landscapes reveal accessible evolutionary paths.
\newblock Nature. 2007;445:383--6.

\bibitem{Loebetal1989}
Loeb DD, Swanstrom R, Everitt L, Manchester M, Stamper SE, Hutchison CA 3rd.
\newblock Complete mutagenesis of the {HIV-1} protease.
\newblock Nature. 1989;340:397--400.

\bibitem{Velazquez-Campoyetal2003}
Velazquez-Campoy A, Muzammil S, Ohtaka H, Sch{\"o}n A, Vega S, Freire E.
\newblock Structural and thermodynamic basis of resistance to {HIV-1} protease
  inhibition: Implications for inhibitor design.
\newblock Curr Drug Targets Infect Disord. 2003;3:311--328.

\bibitem{Louisetal2013}
Louis JM, T{\"o}zs{\'e}r J, Roche J, Mat{\'u}z K, Aniana A, Sayer JM.
\newblock Enhanced stability of monomer fold correlates with extreme drug
  resistance of HIV-1 protease.
\newblock Biochemistry. 2013;52:7678--7688.

\bibitem{Muzammiletal2003}
Muzammil S, Ross P, Freire E.
\newblock A major role for a set of non-active site mutations in the
  development of HIV-1 protease drug resistance.
\newblock Biochemistry. 2003;42:631--638.

\bibitem{Agniswamyetal2012}
Agniswamy J, Shen CH, Aniana A, Sayer JM, Louis JM, Weber IT.
\newblock HIV-1 protease with 20 mutations exhibits extreme resistance to
  clinical inhibitors through coordinated structural rearrangements.
\newblock Biochemistry. 2012;51:2819--2828.

\bibitem{KauffmanWeinberger1989}
Kauffman SA, Weinberger ED.
\newblock The NK model of rugged fitness landscapes and its application to
  maturation of the immune response.
\newblock J Theor Biol. 1989;141:211--45.

\bibitem{Frankeetal2011}
Franke J, Kl{\"o}zer A, de~Visser JAGM, Krug J.
\newblock Evolutionary accessibility of mutational pathways.
\newblock PLoS Comput Biol. 2011;7:e1002134.

\bibitem{DraghiPlotkin2013}
Draghi JA, Plotkin JB.
\newblock Selection biases the prevalence and type of epistasis along adaptive
  trajectories.
\newblock Evolution. 2013;67:3120--31.

\bibitem{Nahumetal2015}
Nahum JR, Godfrey-Smith P, Harding BN, Marcus JH, Carlson-Stevermer J, Kerr B.
\newblock A tortoise-hare pattern seen in adapting structured and unstructured
  populations suggests a rugged fitness landscape in bacteria.
\newblock Proc Natl Acad Sci U S A. 2015;112:7530--5.

\bibitem{Khanetal2011}
Khan AI, Dinh DM, Schneider D, Lenski RE, Cooper TF.
\newblock Negative epistasis between beneficial mutations in an evolving
  bacterial population.
\newblock Science. 2011;332:1193--6.

\bibitem{Chouetal2011}
Chou HH, Chiu HC, Delaney NF, Segr{\`e} D, Marx CJ.
\newblock Diminishing returns epistasis among beneficial mutations decelerates
  adaptation.
\newblock Science. 2011;332:1190--2.

\bibitem{Tokurikietal2012}
Tokuriki N, Jackson CJ, Afriat-Jurnou L, Wyganowski KT, Tang R, Tawfik DS.
\newblock Diminishing returns and tradeoffs constrain the laboratory
  optimization of an enzyme.
\newblock Nat Commun. 2012;3:1257.

\bibitem{Wilkeetal2003}
Wilke CO, Lenski RE, Adami C.
\newblock Compensatory mutations cause excess of antagonistic epistasis in RNA
  secondary structure folding.
\newblock BMC Evol Biol. 2003 Feb;3:3.

\bibitem{Phillips2008}
Phillips PC.
\newblock Epistasis--the essential role of gene interactions in the structure
  and evolution of genetic systems.
\newblock Nat Rev Genet. 2008;9:855--67.

\bibitem{Carteretal2005}
Carter AJR, Hermisson J, Hansen TF.
\newblock The role of epistatic gene interactions in the response to selection
  and the evolution of evolvability.
\newblock Theor Popul Biol. 2005;68:179--96.

\bibitem{Chiottietal2014}
Chiotti KE, Kvitek DJ, Schmidt KH, Koniges G, Schwartz K, Donckels EA, et~al.
\newblock The Valley-of-Death: Reciprocal sign epistasis constrains adaptive
  trajectories in a constant, nutrient limiting environment.
\newblock Genomics. 2014;104:431--7.

\bibitem{Bonhoefferetal2004}
Bonhoeffer S, Chappey C, Parkin NT, Whitcomb JM, Petropoulos CJ.
\newblock Evidence for positive epistasis in HIV-1.
\newblock Science. 2004;306:1547--50.

\bibitem{Arts2012}
Arts EJ.
\newblock Commentary on the role of treatment-related HIV compensatory
  mutations on increasing virulence: New discoveries twenty years since the
  clinical testing of protease inhibitors to block HIV-1 replication.
\newblock BMC Med. 2012;10:114.

\bibitem{Karasovetal2010}
Karasov T, Messer PW, Petrov DA.
\newblock Evidence that adaptation in {{\em Drosophila}} is not limited by
  mutation at single sites.
\newblock PLoS Genet. 2010;6:e1000924.

\bibitem{Rossetal2007}
Ross L, Lim ML, Liao Q, Wine B, Rodriguez AE, Weinberg W, et~al.
\newblock Prevalence of antiretroviral drug resistance and
  resistance-associated mutations in antiretroviral therapy-na{\"\i}ve
  HIV-infected individuals from 40 United States cities.
\newblock HIV Clin Trials. 2007;8:1--8.

\bibitem{Wangetal2011}
Wang D, Hicks CB, Goswami ND, Tafoya E, Ribeiro RM, Cai F, et~al.
\newblock Evolution of drug-resistant viral populations during interruption of
  antiretroviral therapy.
\newblock J Virol. 2011;85:6403--15.

\bibitem{AdamiCerf2000}
Adami C, Cerf NJ.
\newblock Physical complexity of symbolic sequences.
\newblock Physica D. 2000;137:62--69.

\bibitem{Adami2012}
Adami C.
\newblock The use of information theory in evolutionary biology.
\newblock Ann NY Acad Sci. 2012;1256:49--65.

\bibitem{RivoireLeibler2011}
Rivoire O, Leibler S.
\newblock The value of information for populations in varying environments.
\newblock J Stat Phys. 2011;142:1124--1166.

\bibitem{Fano1961}
Fano RM.
\newblock Transmission of Information.
\newblock New York and London: MIT Press and John Wiley \& Sons; 1961.

\bibitem{Basharin1959}
Basharin GP.
\newblock On a statistical estimate for the entropy of a sequence of
  independent random variables.
\newblock Theory Probability Applic. 1959;4:333--337.

\bibitem{Nemenmanetal2002}
Nemenman I, Shafee F, Bialek W.
\newblock Entropy and Inference, revisited.
\newblock In: Adv Neural Inf Process Syst. vol.~14; 2002. p. 471--478.

\bibitem{R-Core-Team2013}
{R Core Team}.
\newblock R: A Language and Environment for Statistical Computing.
\newblock Vienna, Austria: R Foundation for Statistical Computing; 2013.

\end{thebibliography}

\section*{Supporting Information}
% Include only the SI item label in the subsection heading. Use the \nameref{label} command to cite SI items in the text.
\subsection*{S1 Fig.}{\bf  Change in per-site entropies in HIV-1 protease over time.} Average entropy change (compared to 1998) at every position of the HIV-1 protease in the untreated (top panel) and treated (bottom panel) data sets. The size of the circles is proportional to the entropy change, and blue marks an increase while red implies a decrease in entropy at that site, compared to 1998 (the first year in our analysis). Site-specific variation mostly increased across the protein even in the absence of treatment, but decreased at some sites. In the treated data set, the entropy increased at most sites (in particular starting in 2003) while some sites became less entropic.\\
\label{sec:deltas}

\subsection*{S2 Fig.}{\bf  Number of amino acids in each sampled set of 300 sequences for years 1998-2006.} Gaps at the beginning and ends of the protease sequence imply uneven sample size for positions $\ge$15 and $\le$ 90, and thus the ends were truncated for calculation of per-site entropies and pairwise mutual information. Filled circles represent average number of residues in the sampled sets at each protease position and error bars represent unit SD.\\
\label{sec:numbers}

\subsection*{S1 Table}{\bf Protease Inhibitors (PIs) and the years they were approved by FDA.}\\
\label{sec:FDA}

\subsection*{S1 Text} {\bf  Phylogenetic structure of protease sequence data.}\\
\label{sec:tree}

\subsection*{S2 Text} {\bf Sequence Logos for HIV-1 protease sequences.}\\
\label{sec:logo}

\subsection*{S3 Text}{\bf Changes in physicochemical properties at protease positions mirror per-site entropy changes.}\\
\label{sec:chem}
\subsection*{S4 Text}{\bf  Changes in epistatic interactions in a longitudinal study.}\\
\label{sec:long}
\subsection*{S5 Text}{\bf  Information between epistatic pairs in protease of treated and un-treated subjects, for spatially close and distant epistatic residue pairs.}\\
\label{sec:space}
\subsection*{S6 Text}{\bf  Correction for sample-size bias in entropy and mutual information estimates.}\\
\label{sec:bias}
\subsection*{S7 Text}{\bf  Three-loci two-allele model for understanding relationship between epistasis and information}\\
\label{sec:threeloci}

\end{document}